\documentclass[fleqn,10pt]{SelfArx} 

\usepackage[english]{babel} 

\usepackage{lipsum} 

\definecolor{color1}{RGB}{0,0,90} 
\definecolor{color2}{RGB}{0,20,20} 

\usepackage{hyperref} 

\hypersetup{
	hidelinks,
	colorlinks,
	breaklinks=true,
	urlcolor=color2,
	citecolor=color1,
	linkcolor=color1,
	bookmarksopen=false,
	pdftitle={Title},
	pdfauthor={Author},
}

\newcommand{\etal}{\textit{et al}.}

\newcommand{\eg}{\textit{e}.\textit{g}.}

\JournalInfo{Journal Article} 
\Archive{} 

\PaperTitle{Beyond the Screen: Reshaping the Workplace with Virtual and Augmented Reality} 

\Authors{Nuno Verdelho Trindade\textsuperscript{1}*, Alfredo Ferreira\textsuperscript{1}, João Madeiras Pereira\textsuperscript{1}} 
\affiliation{ \textsuperscript{1}\textit{INESC-ID, Instituto Superior Técnico, Universidade de Lisboa, Lisbon, Portugal} \\ \textsuperscript{*}\textit{Corresponding Author}: nuno.verdelho.trindade@tecnico.ulisboa.pt} 

\newcommand{\keywordname}{Keywords} 

\Abstract{Although extended reality technologies have enjoyed an explosion in popularity in recent years, few applications are effectively used outside the entertainment or academic contexts. This work consists of a literature review regarding the effective integration of such technologies in the workplace. It aims to provide an updated view of how they are being used in that context. First, we examine existing research concerning virtual, augmented, and mixed-reality applications. We also analyze which have made their way to the workflows of companies and institutions. Furthermore, we circumscribe the aspects of extended reality technologies that determined this applicability.}

\begin{document}

\maketitle 

\tableofcontents 

\thispagestyle{empty} 


\section{Introduction} \label{introduction} 

\begin{figure}[t]
 \centering
 \includegraphics[width=\the\columnwidth]{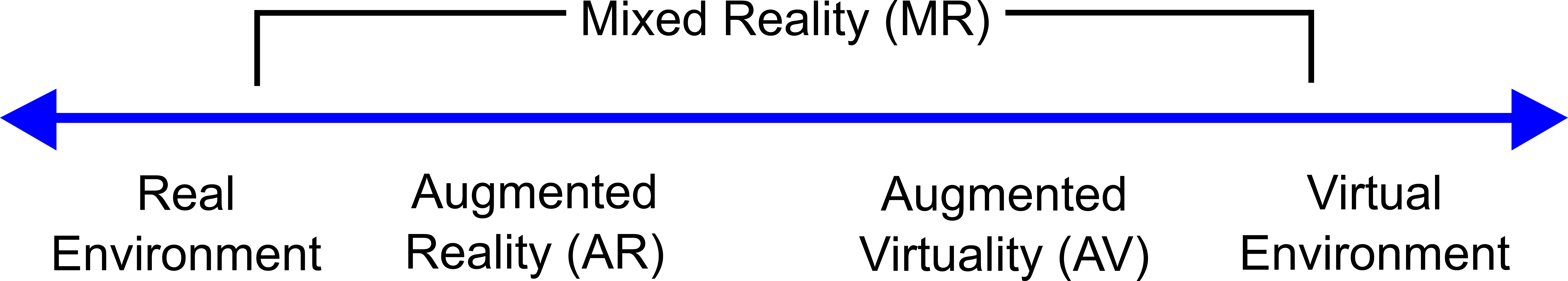}
 \caption{Reality–virtuality continuum~\cite{milgramHumanFactorsConsiderations2006}}
 \label{fig:continuum}
 \end{figure}

Extended reality (XR) refers to the set of real and virtual combined environments such as augmented reality (AR), mixed reality (MR), and virtual reality (VR). The XR spectrum has been conceptualized by Milgram~\cite{milgramHumanFactorsConsiderations2006} in the \textit{reality–virtuality continuum}, which encompasses all possible variations of real and virtual objects (Figure~\ref{fig:continuum}). XR technologies have had considerable growth in popularity in recent years. This growth has been in part due to the evolution of VR and AR/MR techniques. More capable software, application program interfaces, and tracking methods have been developed. We have also seen significant advances in the latest generations of dedicated XR hardware. These advances include increasing the definition and overall quality of cameras and sensors. The growth in speed and efficiency of graphical processors has also been decisive. Another critical factor has been the price cut that occurred in hardware. As a result, XR technologies have had the opportunity to be applied, even in areas where traditional visualization techniques have been long-established. The research work done in the field of XR has shown that this group of technologies can bring tangible advantages in many domains. Nevertheless, it is still the case where only some solutions fall outside the academic or entertainment contexts and have real-world applications.

This work consists of a literature review on the effective integration of XR technologies in the professional environment. It aims to contribute with an updated view of how these technologies are being used in the real world. In that sense, it intends to delimit the aspects that determined its applicability in areas such as engineering and medicine. We start by examining existing research concerning XR applications in different domains. We also analyze which of these applications have made their way to the workflows of companies and organizations. Furthermore, we circumscribe the aspects of XR technologies that determined this applicability. In that scope, we also reflect on future research opportunities.

Relevant XR applications and solutions presented at distinguished conferences in the fields of XR and Human-computer Interaction (HCI) are analyzed. The review includes studies published in the last six years (2015-2020) in the "ACM Symposium on Virtual Reality Software and Technology" (VRST), the "Conference on Human Factors in Computing Systems" (CHI), "Special Interest Group on Computer Graphics and Interactive Techniques Conference" (SIGGRAPH) and the "IEEE Conference on Virtual Reality and 3D User Interfaces" (IEEE VR). We analyzed a total of 4278 articles and selected 164. The methodology adopted for selecting and analyzing the scientific articles is detailed in the following section.

Apart from this introduction, this work presents, in Section~\ref{methodology}, a description of the methodological approach we adopted for the survey of scientific articles included in this literature review. In Section~\ref{literaturereview}, we present relevant research works in the scope of the effective integration of XR technologies in the work environment. This section is organized in the following categories: \nameref{constructionandengineering}, \nameref{transportation}, \nameref{manufacturing}, \nameref{defense}, \nameref{healthcare}, \nameref{sports}, \nameref{education}, and \nameref{other}. The comparison and discussion of the approaches reviewed in the previous section are shown in Section \ref{discussion}. Finally, Section \ref{conclusions} presents the conclusions, including a reflection on possible research directions.

\section{Methodology} \label{methodology}

For the proposed task of getting an updated view regarding the effective integration of XR technologies in the work environment, we decided to focus on relevant distinguished conferences. We analyzed the research work produced during a period of six years (2015-2020) in IEEE VR, VRST, SIGGRAPH, and CHI. These are four "excellent/flagship" conferences in XR and HCI. They are CORE ranked A (IEEE VR and VRST) and A* (CHI and SIGGRAPH) and have featured a high percentage of XR papers since 2015 compared to other conferences with similar rankings. The decision to limit our survey to these four conferences allowed us, on the one hand, to have a term of reference to the evolution of work produced during recent years in applied XR. On the other hand, it also made our survey feasible by limiting it to a meaningful subset.

The choice of papers for our review was based on three complementary criteria. To be selected, an article had to:
\begin{enumerate}
	\item Depict an application of XR to the real world;
	\item Not consist of a purely entertainment-oriented application;
	\item Contemplate the development of a prototype or study based on XR technology.
\end{enumerate}

We queried IEEE and ACM databases for specific XR-related search terms and analyzed the primary studies individually in two phases. First, we opened each study and excluded the papers that did not reference XR technologies in the abstract of keywords. The second phase of the selection consisted of doing a more thorough analysis. Based on the previously mentioned selection criteria, we excluded the papers that did not depict an application of XR to the real world. This manual process, although time-consuming, allowed for a more refined and reliable analysis of the research papers. From a total of 4278 articles (1092 from IEEE VR, 190 from VRST, 1370 from SIGGRAPH, and 2996 from CHI), we selected 171.

\begin{figure}[b!]
\centering
\includegraphics[width=\the\columnwidth]{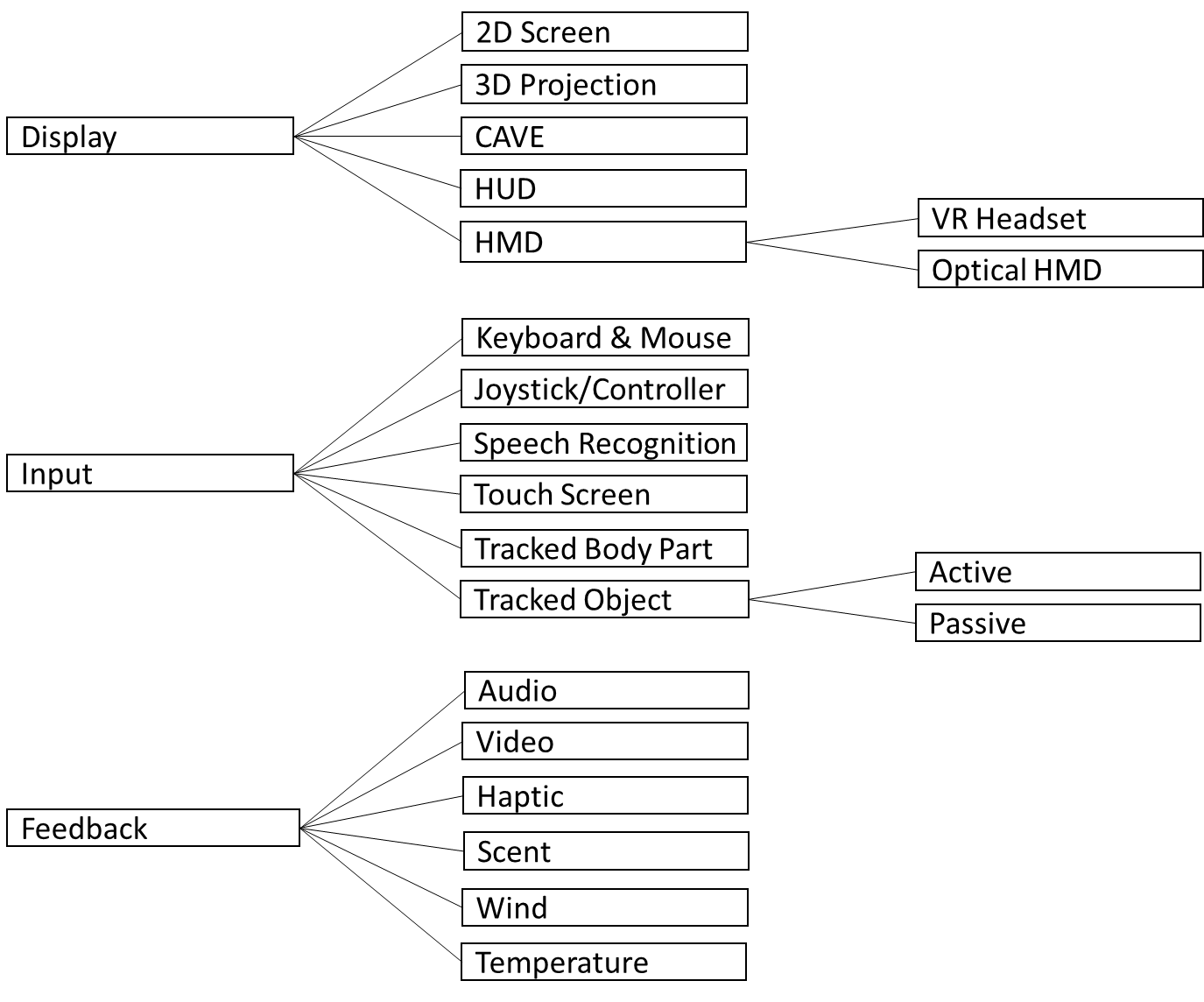}
\caption{Input and output taxonomy used in the comparison between approaches}
\label{fig:taxonomy_01}
\end{figure}

A comparison between the different approaches is done in Section \ref{discussion}. This comparison encompassed two types of assessment. First, we categorized the type of display, input, and feedback aspects of the solutions presented in the analyzed works. This process was carried out using the taxonomy presented in Figure~\ref{fig:taxonomy_01}. Next, we used the Global Industry Classification Standard (GICS) methodology~\cite{Gics2020} to frame the different solutions in sectors and industries.

\section{Literature Review Categories} \label{literaturereview}

This section presents several practical works regarding applying VR, AR, MR, and Augmented Virtuality (AUV) technologies to the real world. We explore applications and solutions that have been developed in different fields. In the area of \nameref{constructionandengineering}, we present solutions in the visual inspection of large civil engineering structures, Building Information Modeling (BIM), Traffic Engineering, municipal infrastructure works, architecture, interior design, urban projects, and occupational health and safety. Regarding \nameref{transportation} \& Logistics, XR solutions are addressed in assisted car driving, interfaces in see-through cockpits, driving simulation, road safety, automotive production, aircraft in-flight interfaces, and maritime travels. We also present examples of applications to scuba diving simulation and underwater work. For the \nameref{manufacturing} industry, practical examples are analyzed in industrial layout planning, industrial robot manipulation, and manual manufacture, among others.

Regarding the \nameref{defense} sector, solutions for Armed Forces, Law Enforcement, and Fire Protection are addressed. These include combat vehicle coordination, military weapons handling, situational awareness, de-escalation training, and aerial/ terrestrial firefighting training. Furthermore, we analyze XR examples of microgravity locomotion in space missions and the design and engineering of space vehicles. In the \nameref{healthcare} area, we cover applications such as surgical intervention assistance and training, medical imaging, design of surgical instruments, nursing and medical training, prosthetics, rehabilitation, exposure therapy, and accessibility. Applications to \nameref{sports} are presented for basketball, baseball, golf, jogging, and motor learning. In the cultural heritage domain, we analyze interactive virtual environments that simulate actual prehistoric archaeological sites and also an immersive excavation VR simulation. Regarding \nameref{education}, we address applications for architecture undergraduates to build design models, collaborative automotive assembly tasks, teaching history in middle school, or physics education applications.

\subsection{Engineering Consulting Services} \label{constructionandengineering}

The Architecture, Engineering, and Construction (AEC) sector has been on the front line concerning the early adoption of emerging, innovative and environmental technologies~\cite{verdelhotrindadeEnvironmentalConcreteNew2009}. In that sense, XR applications have been developed for architectural design, urban planning, construction management, and structural health monitoring, among others.

Regarding the visual inspection of large civil engineering structures, Veronez \etal~\cite{8798295} created \textit{Imspector}, an immersive system for inspecting bridges and viaducts. The system uses mapping data gathered by Unmanned Aerial Vehicles (UAVs) with laser scanning to generate 3D models that can be integrated into virtual environments. These immersive environments can be explored by the Structural Engineer using VR headsets. Such systems allow the inspection to be made off-site by specialized professionals, while less specialized staff can handle data gathering on-site. The use of VR, in this case, also has the potential of hindering one of the main difficulties when inspecting large structures, which is the difficulty in physically accessing specific locations~\cite{verdelhotrindadeDamARAugmentedReality2019}.

BIM has been a constant buzzword in the AEC sector for the last decade. In the context of BIM, Raimbaud \etal~\cite{8446288} created a way of adapting BIM data for virtual environments. The devised methodology uses filtering techniques to select which segments of the BIM raw data help create 3D elements for VR visualizations. Raimbaud \etal~\cite{8797784} also addressed using MR to verify construction accuracy. They developed a system that allows a Civil or Structural Engineer to compare what has been effectively built in the construction site with what was originally planned in the project. This comparison is made by superimposing BIM models with real images captured by UAVs.

In the area of Traffic Engineering, Uhr \etal~\cite{8446141} built a decision support system for traffic planning that uses a hybrid VR setup. The system combines interactive display technologies, including an interactive tabletop and a Cave Automatic Virtual Environment (CAVE) projection. Engineers can collaboratively plan traffic interventions around the touch tabletop in coordination meetings. Veronez \etal~\cite{8446207} developed a VR system to assist Civil Engineers in road design. Using a VR headset and a steering wheel for a PC, the designer can assess how driving through a projected road section will feel.

In the scope of municipal infrastructure works, Côté \etal~\cite{8446545} developed an AR method for the augmentation of subsurface utility pipes. While this type of solution, which allows the user to have a kind of "X-ray vision" of what is under his feet, has been extensively studied in the past~\cite{Schall2009, Behzadan2009, 4637332}, this new method promises increased accuracy. The authors propose using pre-captured photorealistic 3D meshes of the roads as a reference for the alignment of the pipe network. This reference prevents misalignment problems, where the pipes appear to float above the road due to variable topography.

Ergün \etal~\cite{8798180} explored the use of VR and MR, together with already-existing design tools for architecture. They developed a process that allows the seamless integration between BIM tools (\eg, Autodesk Revit\footnote{Autodesk Revit: https://www.autodesk.com/products/revit/overview}) and 3D graphical engines typically used for XR visualizations (\eg, Unity\footnote{Unity: https://unity.com/}). Such a process fosters a continuous bidirectional workflow that allows the expansion of the capabilities of BIM. Boustila \etal~\cite{10.1145/2821592.2821595} studied how distances are perceived in virtual environments in an architectural framework. Based on the results, they present guidelines for setting up architectural project review tools. Also, in that scope,  Boustila \etal~\cite{7504702} addressed the perception of the near surrounding ground in virtual environments. Kán \etal~\cite{8448291} devised a method for automatic interior design using VR environments. A procedural approach was used, which allows a fast generation of furniture arrangements in interior scenes. Moraes \etal~\cite{7892347} focused on the automation of 3D modeling for integration in virtual environments, namely in generating 3D models of hydroelectric power plants. In the framework of urban projects, Vigier \etal~\cite{7223417} studied climate perception in virtual urban environments. They focused on the role of visual cues in the user's perception of seasons, time of day, and temperature.

VR has also been used in the context of occupational health and safety. Di Loreto \etal~\cite{8448292} created  "\textit{WoaH}", a VR work-at-height simulator. The system can be used to train workers in work-at-height engineering operations, like constructing or inspecting bridges, high-rise buildings, or dams. \textit{WoaH} uses an HMD with a real ladder synchronized in position with a virtual one in the VR environment. On the other hand, Chardonnet~\etal~\cite{8446395} devised a VR system to detect Acrophobia (the irrational fear of heights) in work-at-height situations. The solution subjects the user to a high-fidelity VR simulation that triggers the fear of heights. Nickel \etal~\cite{7223457} created a VR simulation for occupational safety and health assessment during hydraulic machinery design. The system simulates the operation of river locks and is intended for use by risk assessment inspectors. The contextualization of construction accident reports was addressed by Peña \etal~\cite{7892340}. The system allows users to explore a construction site in a VR environment, interacting and learning more about a particular accident. Shaw \etal~\cite{10.1145/3290605.3300856} built a VR setup for the fire safety assessment of buildings. They associated the traditional audiovisual (AV) experience an HMD provides with thermal and olfactory stimuli. Fire is simulated by a regulable intensity infrared heater that turns on when the user is in the proximity of flames in the virtual environment. The smell is simulated using a dedicated scent dispenser with a dispersion fan. These additional stimuli allow a more immersive experience that can better capture user behavior in a set of fire evacuations.

\subsection{Road, Rail, Airlines and Marine Transportation} \label{transportation}

Transportation and logistic networks are an essential part of our continuously connected world. Massive quantities of people and goods circulate by land, air, and water daily. In that sense, even minor optimizations in the transportation processes can result in considerable savings both in time and money. XR technologies can be essential in such optimization, and multiple applications have been studied for the automotive, maritime, and aviation sectors.

AR car driving using semantic geo-registration was addressed by Chiu \etal~\cite{8447560}. They developed a method that uses 3D georeferenced data to generate a rendered depth map. This depth map is used to blend, in a realistic manner, 3D graphics with the real world. An MR solution that addresses the future emergence of transparent cockpits in the automotive industry was developed by Lindemann \etal~\cite{7892298}. In this interface concept, camera images from the outside are projected onto the car interior. By using diminished reality, the system allows drivers to perceive otherwise occluded objects in the real environment through the car body. Lindemann \etal~\cite{8798069} analyzed the implications of driving with see-through cockpits in narrow-space overtaking scenarios. Lindemann \etal~\cite{8797957} also examined the effects of inaccurate head tracking on drivers of vehicles with transparent cockpit projections.

To more accurately reproduce the rich sensations of real driving, Goedicke \etal~\cite{10.1145/3173574.3173739} devised "\textit{VR-OOM}" the first on-road VR driving simulator. Unlike a typical VR driving setup where the user is static, \textit{VR-OOM} implements driving simulation in a vehicle as it travels on the road. The user, wearing a VR headset, sits on the passenger seat where a gaming steering wheel is installed. During the simulation, a trained driver sits in the driver's seat of the real car and mimics the user's steering behavior. The improvement of driving skills using VR-powered solutions was addressed by Lang \etal~\cite{8448290}. They used a driving controller associated with an eye-tracking VR headset, which allows for collecting both the user's driving and eye gaze data. Likewise, Sportillo \etal~\cite{8446226} used VR to implement a training and learning tool to familiarize drivers with automated vehicles. In particular, the system aims at improving take-over performance. The work of Ju \etal~\cite{7504690} can also help improve professional drivers' behavior on the road. Their study was focused on how drivers make decisions in challenging situations, like traffic accidents, using VR. Hoesch \etal~\cite{8446240} explored the relationship between visual attention and motion sickness in VR driving simulators. In road safety, Kim \etal~\cite{7504725} designed a novel interface for pedestrian collision warning aimed at automotive AR head-up displays (HUD). The system casts what the authors describe as "virtual shadows" of pedestrians and other obstacles in the AR environment. The shadows turn red to warn the driver of an urgent situation that requires his immediate reaction (\eg, the car is in a collision course with a pedestrian).

An AR version of the popular "tunnel-in-the-sky" interface for aircraft cockpits was devised by Gorbunov \etal~\cite{7223449}. The traditional "tunnel-in-the-sky" interface allows pilots to visualize 3D navigation paths on a flat screen. The AR version requires the pilot to use a lightweight AR headset and offers stereoscopic vision. In the scope of maritime travels, Stevens \etal~\cite{8797800} studied the use of VR to reduce seasickness. They developed a solution aimed at hindering motion sickness on ship crews, which moves the surroundings in the VR environment to match vessel motion. Exploring harsh or remote environments without being physically there is also one of the virtues of VR. On the other hand, using AR and MR can be an important asset in environments with low visibility, where the user is wearing a protective suit or where using traditional interaction devices would be challenging. An example of such a setting is underwater work.

Hatsushika \etal~\cite{8798052} developed "\textit{SCUBA VR}", which allows subsea workers to experience arbitrary underwater conditions in a VR environment. The system has the unique feature of being able to be operated in a pool or shallow water using a waterproof VR headset. To test the additional limitations on tracking accuracy when underwater, Costa \etal~\cite{7892281} evaluated a set of commercial consumer tracking systems. The results suggest that magnetic tracking systems are the most suitable for underwater VR applications. On the other hand, Yamashita \etal~\cite{8446490} devised a water flow measurement technology aimed at spatial user interaction in underwater immersive VR environments. They use transparent tracer particles that are scattered in the fluid and whose movement can be tracked by cameras using polarization-based technologies.

\subsection{Industrial Machinery} \label{manufacturing}

The manufacturing industry is constantly searching for new ways to improve production processes. Advanced automation technologies, including robotic systems and machine learning, are becoming the norm on the industrial floor. The application of XR technology has been studied in such contexts as designing industrial facilities or planning production lines.

In the framework of factory design, Gebhardt \etal~\cite{7223355} analyzed the use of XR for industrial layout planning. They created "\textit{flapAssist}", a VR factory layout planning assistant. The tool offers designers virtual walkthroughs of the planned factory and the possibility of visualizing information regarding the dynamics on the industrial floor. The project "\textit{Factory-in-a-day}", which aims to optimize hybrid robot-human production lines installation, was presented by Aschenbrenner \etal~\cite{8446533}. The project uses AR technologies to allow engineers to collaboratively decide where the different components of the production line should be installed.

The use of AR immersive telepresence for robot manipulation in industrial manufacturing was addressed by Peppoloni \etal~\cite{10.1145/2821592.2821620}. They developed a system that allows the user to control the robot using a wearable device. This device allows the movements and muscle contractions of the operator to translate into actions of the robot arms. In the scope of AR-based assistance systems, Renner \etal~\cite{8446127} addressed user support in search tasks by guiding their attention toward relevant targets. They developed a prototype for assembly environments to reduce wrongly picked items or false placements. This reduction is achieved using a 3D path guiding that directs the user towards the target.

Industrial maintenance has also been the target of multiple XR solutions. Winther \etal~\cite{9089646} worked with the Danish manufacturer Grundfos to develop a VR training simulation. This solution is directed at sequential pump maintenance task learning without interfering with industrial systems' operation. Burova \etal~\cite{10.1145/3313831.3376405} focused on safety aspects in industrial maintenance activities. They used VR simulations to prototype industrial AR solutions that can be used on-site, namely in the elevator maintenance industry.

AR and VR applications have also been researched for manual manufacture. Freeman \etal~\cite{7223458} devised an AR tool for assistance during composite material fabrication. The solution guides the worker during this process by overlaying reality virtual guidelines regarding the geometry of the composite element that is being fabricated. Xie \etal~\cite{7223419}, on the other hand, created "\textit{Onew360}", a real-time manual welding training system based on VR. Büttner \etal~\cite{10.1145/3313831.3376720} researched efficient on-the-job training with the help of assistive AR systems. They used projection-based AR in the scope of industrial assembly work to decrease training time and improve how well a learned task is remembered.

\subsection{Aerospace \& Defense} \label{defense}

There are few areas of society where adequate tools, thoughtful planning, careful design, and efficient execution are more critical than in the defense sector. Because many human lives depend on it, governments spare no expense in equipping the Armed Forces, Law Enforcement, and Fire Protection with the most proven and effective technologies. Consequently, the private defense industry is one of the most profitable. Active research \& development of defense applications using the latest technologies is encouraged to get an edge over competitors. In the framework of XR applications, the military has pioneered AR, \eg, in combat aircraft HUDs, and VR, \eg, for training tank and armored vehicle crews. However, XR technology applications have also been researched in other scopes, including firefighting, law enforcement, and space exploration.

Khooshabeh \etal~\cite{7892312} designed "\textit{TALK-ON}", an MR prototype aimed at simulating the role of a tank platoon leader. The leaders are immersed in a virtual battlefield simulation, where they can coordinate operations with a virtual crew and other virtual tank leaders. Taupiac \etal~\cite{8797854} devised a VR system for training rifle infrared (IR) sight calibration. Using a VR headset and a rifle mockup with integrated VR controllers, the soldier is put in a virtual environment that simulates a shooting range. Clements \etal~\cite{8446068} studied the cognitive and neural mechanisms underlying motor skill learning in simulated VR marksmanship training. They used a CAVE-like system and a mockup rifle to subject a group of twenty marksmen to a three-day VR training regimen in a virtual shooting range. Brandão \etal~\cite{7892294} conceptualize using AR technologies to improve dismounted operators' situational awareness. In that scope, they point out essential visual elements that should be considered in AR systems for military and law enforcement applications. These include the superimposition of navigation data and the position of allied forces and enemies to the real world. Hughes \etal~\cite{7504713} applied AUV technologies to law enforcement de-escalation training. In this virtual environment, law enforcement officers are subjected to a stressful scenario involving a highly agitated individual. The system supports verbal communication with the virtual characters. The system also detects physical actions like reaching for the gun holster.

Suhail \etal~\cite{8798280} devised a VR system that simulates the operation of firetruck pumps in fire protection settings. The solution was created for first responders training and associates using VR headsets with a physical firetruck pump panel. This combination introduces passive haptics in the virtual environment, which increases the immersiveness and sense of presence. In the context of VR aerial firefighting training, Clifford \etal~\cite{8797889} analyzed the effect of different types of displays on situation awareness. Tests in air attack supervision simulations allowed the researchers to conclude that better situation awareness was achievable with immersive displays. Clifford \etal~\cite{8446139} addressed the creation of stressful decision-making environments for aerial firefighter training in VR. To produce a realistic training environment, the researchers developed a multi-sensory VR flight simulator system that combines visual and auditory with tactile stimuli. It uses a 270 $^{\circ}$ cylindrical display system with three aligned projectors, surround sound, and vibration induced through an actuator under the seat. Jeon \etal~\cite{10.1145/3359996.3364268} identified aspects of firefighting scenarios that should be incorporated in VR firefighting-training-systems to make them more realistic. Mossel \etal~\cite{7892324} developed the "\textit{VROnSite}" platform that enables immersive training of first responder squad leaders in VR.

An underwater VR system that simulates jetpack locomotion outside the International Space Station (ISS) was developed by Sinnott~\etal~\cite{10.1145/3359996.3364272}. The system aims to emulate conditions experienced by astronauts during extra-vehicular activities (EVA). For their part, Ferrer \etal~\cite{8798253} studied human visual exploratory activity in a microgravity environment with the aid of VR. They equipped subjects with VR headsets and analyzed, during parabolic flights, their reaction to a virtual representation of space and the ISS. The reduction of the adverse effects of confinement and monotony in long-duration space and underwater missions using VR experiences was addressed by Solignac \etal~\cite{7223408}. They developed a prototype that immerses the crew members in visually rich virtual landscapes with distant horizons, plants, animals, and climate. The locomotion through the virtual world is done using an ergometer. Soccini \etal~\cite{7223463} applied VR metaphors to the design and engineering of space vehicles. They developed two applications that allow engineers to perform aerothermodynamics analysis in a VR environment.

\subsection{Health Care} \label{healthcare}

The health sector has been one of the primary consumers of advanced and innovative visualization techniques. This interest comes from the constant need to understand better and systematize the processes inherent to human anatomy, physiology, and general well-being. In that scope, XR technology has been extensively researched in different healthcare domains. AR and VR solutions have been developed for professional surgical applications, medical education and training, rehabilitation, accessibility, and even veterinary medicine.

Eagleson \etal~\cite{7223349} developed a prototype for planning surgical interventions. The "\textit{NeuroTable}" system allows real-time collaboration between neurosurgeons over an augmented tabletop during intervention planning. The surgeons use a haptic stylus to interact with the visualization of the anatomical structures. In the context of neurosurgical planning, Fiederer \etal~\cite{8797710} created a framework for VR neurosurgery planning that can provide neurosurgeons, wearing VR headsets, with the tools to inspect high-accuracy 3D models of the patients. Rapetti \etal~\cite{10.1145/3139131.3139162} devised a surgical navigation system for performing targeted prostate biopsies. The solution allows the continuous tracking of the biopsy needle in relation to the patient's anatomy during the procedure. Using a stereoscopic 3D monitor, the surgeon can guide the virtual representation of the tracked needle through several anatomical layers toward the biopsy target.

In the "\textit{VRRRRoom}" project, Sousa \etal~\cite{10.1145/3025453.3025566} created a VR solution to assist radiologists in unsuitable ambient conditions. \textit{VRRRRoom} combines a VR headset with an interactive surface to offer the radiologist an immersive environment free of ambient interferences. Ard \etal~\cite{7892381} created the visualization suite "Neuro Imaging in Virtual Reality" (NIVR). The suite is directed at representing neuroimaging information in a VR environment. Wearing a VR headset, the user can explore a volumetric texture constructed from an MRI scan. Duncan \etal~\cite{7892382} created a VR-assisted system for improving the efficiency of algorithmic segmentation of MRI scans. Lorenz \etal~\cite{8798144} analyzed the application of VR in the development of surgical instruments. They studied the use of VR to obtain feedback from medical staff in the initial phase of developing a surgical aspiration/irrigation instrument.

A VR trainer for endotracheal intubation training, named "\textit{AirwayVR}", was developed by Rajeswaran \etal~\cite{8798249, 8446075, 8797998}. The procedure simulation is organized into smaller learning objectives so that the user can learn intubation step-by-step. A similar system, but directed at neonatal endotracheal intubation training, was presented by Xiao \etal~\cite{9089542}. This system uses a haptic pen to improve feedback and guidance during the training procedure. Ricca \etal~\cite{7892259} designed different navigation techniques for needle insertion in a virtual biopsy trainer. Qian \etal~\cite{10.1145/2821592.2821599} addressed laparoscopic surgery simulation using VR. The interaction with the virtual environment is done using two haptic styli (\textit{Omni Phantom}\footnote{Phantom Omni: https://www.immersion.fr/en/phantom-touch/}), which can accurately emulate real laparoscopic instruments.
Similarly, Li \etal~\cite{9089445} produced a laparoscopy procedure training solution with haptic feedback. They focused on improving usability and presence and evaluated the system with surgeons and surgical trainees in a real-world scenario. Trejo \etal~\cite{8446310} addressed the application, in VR medical simulations, of force models that estimate soft-tissue biomechanical responses. Medina-Papaqayo \etal~\cite{8446504} developed a VR solution for simulating intraosseous access in newborns. It simulates the insertion of a needle beneath leg tissues to the bone. During the simulation, the trainee is guided through the steps needed to perform the technique. The user, wearing a VR headset, interacts with the system through a haptic stylus.

Kaluschke \etal~\cite{8446462, 8446370} created a VR hip prosthesis implantation simulator incorporating haptic feedback. In order to simulate the forces involved in the milling process, an industrial robotic arm was used. During the simulated procedure, the trainees wearing VR headsets hold the extremity of the robotic arm. They can then feel the vibration and other types of feedback expected in the actual procedure. Liyanage \etal~\cite{8798159} devised a VR-based solution for Cardiopulmonary Resuscitation (CPR) training. It uses a physical, mechanical manikin for haptic feedback and a depth camera-type device to track the movement of the hands during the resuscitation maneuver. Si \etal~\cite{8446450} presented an AR interactive environment for neurosurgical training. This environment uses a 3D-printed skull as a base for the virtual augmentation. The trainee uses a pair of haptic styli, which can simulate the sensation produced by making incisions in the brain mass. The visualization is done using an AR headset.

Also, for neurosurgical training, de Ribaupierre \etal~\cite{7223338} developed an AR tool that simulates the placement of an external ventricular drain in the brain. The system associates a mockup head with a haptic stylus to produce feedback. The AR visualization is done using a tablet. Mathur~\cite{7223437} created a low-cost VR solution for general medical training. The researcher devised a virtual environment where trainees can practice identifying body parts or performing incisions in virtual patients. The interaction with the immersive environment is done with a game controller. Daher~\cite{7892354} analyzed and compared the use of conventional AR and augmented optical see-through in simulators for medical applications. For that purpose, the researcher developed an interactive touch-sensitive physical-virtual head that reacts verbally and non-verbally to touch.

Wijewickrema \etal~\cite{10.1145/2993369.2993397} developed an automated guidance process aimed at replacing expert supervision during surgical VR training. The method provides trainees with step-by-step advice regarding a specific type of surgery. Todsen \etal~\cite{8446469} used stereoscopic 360$^{\circ}$ VR video to give medical students a realistic experience of the operating room. Schild \etal~\cite{8446160} created a multi-user VR solution for collaborative medical training. The system allows two trainees, wearing VR headsets, to interact in the same virtual training session while being supervised by a trainer. Sankaran \etal~\cite{8798089} developed an interactive MR prototype to enhance simulated medical training by accelerating clinical exposure for novice medical students. The system uses 360$^{\circ}$ video recording to subject the trainee to a clinical encounter from a first-person perspective. The system uses a VR headset, and the interaction is carried out with VR controllers. Dey \etal~\cite{8797840} used VR technologies to create a cognitively adaptive system for medical training. The system measures real-time alpha activity in the brain using EEG sensors. The difficulty of the tasks in the VR training simulation is then dynamically adapted according to EEG measure values. The evaluation of EEG information of users while experiencing a virtual environment was addressed by Tauscher \etal~\cite{8797858}. They modified a VR headset with EEG sensors to collect EEG data.

Azimi \etal~\cite{8446583} addressed the use of MR for the training of caregivers. Wearing an AR headset, the users are guided in tasks like needle chest decompression and installing a direct intravenous line. The augmented elements are superimposed on a physical patient care manikin, allowing trainees to practice using real medical instruments. Dong \etal~\cite{8797918} developed "\textit{cryoVR}", a passive haptic feedback VR system for training in bio-sample preparation. The VR environment mimics a real lab, and the virtual objects are superimposed on existing physical elements in the real world. Choi~\cite{8797741} created a VR wound care training system for clinical nursing education. The solution allows the trainee to experience the steps of changing a simple wound dressing in an immersive environment. Tani \etal~\cite{8797761} devised a VR simulation that allows users to experience bidirectional infections physically. The system represents the cloud of pathogens spreading through the air and infecting the other parts. If the users get infected, they experience vibration in the chest induced by vibrotactile transducers. Inks \etal~\cite{8446502} developed a solution that introduces virtual avatars for social interaction in medication trials. These virtual characters are responsible for explaining the experiment, medication, and risks involved. The method aims to provide a consistent and standardized interaction between participants and experimenters. Voinescu \etal~\cite{8798191} created the "Nesplora Aquarium", a VR solution for neuropsychological assessment of the attention of adult patients. The participants are evaluated by how they respond to a series of visual and auditory stimuli in the VR environment. In the field of veterinary medicine, Seo \etal~\cite{7892345} developed "\textit{Anatomy Builder VR}". This VR system is directed at learning canine anatomy.

XR technologies have also been extensively researched for physical rehabilitation purposes. Debarba \etal~\cite{8446368} created an AR tool for human motion analysis that allows healthcare professionals to visualize joint movements. It offers a kind of "X-ray vision" that shows, superimposed to the patient, an anatomical representation of the joint structures. Spicer \etal~\cite{7892338} developed a low-cost neurofeedback system called "\textit{REINVENT}". The system is directed at rehabilitation, using VR environments, of severe motor-impaired patients as a result of a stroke. \textit{REINVENT} integrates EEG and electromyography (EMG) sensors and can provide neurofeedback to the VR simulation. This direct connection between brain activity and the simulation of movement has been known to encourage greater use of limbs in stroke patients~\cite{ballester2015}. Phelan \etal~\cite{7223441} explored prosthetic limb training using VR. The system uses a depth camera and a wearable gesture recognition armband to track the position of the arms. Heidari \etal~\cite{8798321} created a VR solution that incorporates human-limb motion tracking and visualization to be used to design cooperative human-robot systems. Covarrubias \etal~\cite{10.1145/2821592.2821619} devised a VR immersive system for upper-limb rehabilitation. The VR environment stimulates the user to perform a series of exercises designed explicitly for upper-limb recovery. The system uses a depth camera to track hand movements and to detect gestures. Marquardt \etal~\cite{8446553} studied the use of multisensory VR environments in the supportive treatment of different anxiety disorders. They devised a setup that includes speakers, fans, olfactory devices, and tracking devices. The user, wearing a VR headset, is subjected to a range of multisensory stimuli that can be directed to trigger different kinds of emotional responses.

The use of VR technologies for chronic pain management was addressed by Gromala \etal~\cite{10.1145/2702123.2702344}. The "\textit{Virtual Meditative Walk}" (VMW) system was designed for patients to learn mindfulness-based stress reduction (MBSR) meditation. The VMW  transports users to a peaceful, non-distracting, and safe virtual environment that reduces chronic pain. The system includes galvanic skin response (GSR) sensors that measure the arousal levels of the patient. The elements of the virtual environment are dynamically adapted according to these levels. VR applications have also been researched in alcohol use disorder therapy. Mostajeran \etal~\cite{8797817} adopted a gamification approach based on behavioral therapy methods that use relapse-risky environments. Saito \etal~\cite{8798081} explored the use of AR for reducing Phantom Limb Pain (PLP) syndrome, which manifests in patients who recently lost a limb. Using an AR headset, the user is subjected to mirror therapy. In this treatment, the user is shown, in an augmented environment, a virtual limb that occupies the place of the lost limb. The existing hand and arm movements are then mirrored to the virtual arm. Infrared cameras detect the spatial locations and actions of the intact arm and fingers. Hurd \etal~\cite{8797997} addressed the treatment of Amblyopia (weaker vision in one eye) through VR. The system consists of a video game paired with a physical monocular, which promotes correcting the neurological connection between the lazy eye and the mind.

Gait rehabilitation using VR was addressed by Hamzeheinejad~\cite{8797872, 8446125, 8797763}. This type of rehabilitation consists of subjecting patients suffering from post-stroke motor impairments to repetitive practices using a robot-assisted gait device. The VR method immerses the user in calming virtual environments, promoting escapism to nature. As the patients walk in the gait device, wearing a VR headset, they feel like they are walking through a forest or a beach. This process tends to suppress the clinical context. Kern \etal~\cite{8797828} also studied the application of VR environments in gait rehabilitation but from a gamification perspective. Bekele \etal~\cite{7504695} devised a therapeutic VR solution for children with Autism Spectrum Disorders (ASD). The system promotes interaction between the children and virtual characters in a virtual social environment. The system includes an eye tracker, an EEG monitor, and biosensors to measure peripheral electrophysiological signals. The VR simulation is dynamically adapted depending on the sensors' real-time measurements to address the child's emotional state. Adjorlu \etal~\cite{8798032}, on the other hand, devised categories of skills that could be taught efficiently to children and adolescents with ASD using VR. Multiple VR applications have also been researched in the physical accessibility of persons with disabilities. Maidenbaum \etal~\cite{7223435} applied VR technologies in the training of users who are blind, in the effective use of a cane.

Choo \etal~\cite{10.1145/3290605.3300605} devised an AUV solution directed at mobile app accessibility design. The system allows developers wearing a VR headset to experience how their mobile apps appear to individuals with visual disabilities. Krösl \etal~\cite{8798239} used VR technology to simulate vision impairments. Architects and lighting designers can use this tool to understand how interior and urban spaces can be improved. Chowdhury \etal~\cite{8446132, 8446146} used VR to simulate the opposite effect. They focused on designing VR environments, allowing people with disabilities to perform tasks more efficiently than in the real world. Kanno \etal~\cite{8446143} created an AR mobile application to help individuals with early Alzheimer's Disease identify objects and people. Renner \etal~\cite{8446292} developed a simulated AR application (using VR) to help impaired and older adults pick and assemble. Tabbaa \etal~\cite{10.1145/3290605.3300466} developed a VR solution to provide accessible experiences for people with dementia in locked psychiatric hospitals.

Tarng \etal~\cite{8798266} focused on decoding brain behavior during interaction in virtual environments. They tried to correlate EEG patterns with the performance of specific tasks in haptic VR simulations. Nakano \etal~\cite{8798336} designed an AR system that uses video see-through, which can change the visual appearance of food in real time. The effects of cybersickness on persons with multiple sclerosis were addressed by Arafat \etal~\cite{8446194, 10.1145/2993369.2993383}. On the other hand, Koilias \etal~\cite{8798084} studied how a VR trip with a self-driving car could affect participants' anxiety. Ferdous \etal~\cite{8446488} analyzed how balance-impairments induced by diseases like multiple sclerosis could contribute to postural instability in virtual environments.

\subsection{Sports/Leisure Products and Facilities} \label{sports}

The sports industry can benefit significantly from XR applications. Professional athletes can improve both their performance and technique in VR environments. The training in VR can be achieved with the help of virtual coaches or by mimicking high-performance athletes in the virtual environment. On the other hand, AR and MR can be used, \eg, to superimpose virtual running trajectories and obstacles to the real world. Because of such possibilities, XR applications have been devised in sports as diverse as athletics, basketball, baseball, golf, and even climbing.

Tsai \etal~\cite{8798309} developed a basketball offensive decision-making VR training system. Wearing a VR headset, the player is immersed in a virtual basketball court. The position and movement of the player in the court are tracked using an inertial measurement unit (IMU) based motion capture suit. The system uses deep learning and can recognize if the specific offensive maneuver performed by the user was the most correct. Zou \etal~\cite{8798041} created a VR-based baseball batting training solution. The system aims to improve the batting performance of professional baseball players. Using a VR headset and a real bat, the player can practice baseball swings in a virtual environment that simulates a traditional batting center. Isogawa \etal~\cite{8446073} studied the requirements needed for replacing real with virtual environments in the scope of baseball training. They compared the factors contributing to the batter's reaction in VR environments with distinct configurations.
On the other hand, Ryge \etal~\cite{7892328} addressed incorporating high-fidelity haptic feedback to simulate the impact between a virtual baseball bat and a ball. They built a prototype that allows the user, wearing a VR headset, to experience the physical feeling of hitting a ball with a bat. This stimulus was achieved by combining a controller with a haptic actuator with an additional vibrotactile transducer controlled by audio signals.

Godse \etal~\cite{8798026} investigated using VR environments in golf training. They manipulated the properties of objects in the virtual environment, namely the size of the golf balls. They then evaluated how that size change influenced the perception of a golf-putting task. The use of VR has also been researched in the scope of outdoor climbing sports. Tiator \etal~\cite{8446506} developed "\textit{Cliffhanger-VR}", a VR solution directed at making climbing safer. Wearing a VR headset, the users are transported to a situation where they are suspended on a high cliff. The position of the virtual cliff wall is matched with a real low-height climbing wall that the users can climb. Hamada \etal~\cite{7892371} devised an AR jogging companion to augment users' daily running experience. The system shows a virtual runner in an AR environment to the user wearing a see-through HMD. This virtual character runs at a preconfigured pace, and its motion is automatically synchronized with the user's movement. Shimizu \etal~\cite{8798227} created a VR training solution for improving spatial awareness in team games. For that, the system allows the user, wearing a VR HMD, to alternate between a bird's-eye view of the sports ground and a first-person perspective of the players and the ball. Waltemate \etal~\cite{10.1145/2821592.2821607} developed a low-latency VR environment for motor learning. The system uses a CAVE environment to display a virtual fitness room. The user, wearing a pair of tracked 3D glasses, has his movement captured by an optical motion tracking system mounted at the top and the sides of the CAVE.

As time passes, both physical artifacts and intangible resources from past generations acquire an increasingly delicate and ethereal condition. In that sense, digital technologies can be invaluable in conserving historical and cultural assets. In domains like archaeology, XR solutions can be used as tools to visualize and better understand the past.

Borba \etal~\cite{7892380, 7892326, 7504785} developed "\textit{ArcheoVR}", an immersive and interactive virtual environment that simulates an actual prehistoric archaeological site. This virtual environment can be explored using a VR headset and controllers. The deployment of VR in a museological setting was demonstrated by Rehnberg \etal~\cite{10.1145/3388530.3412518}. They addressed the safety and security issues while implementing an on-site VR system at the Fort Worth Museum of Science and History.

To provide embodied knowledge and situated experience, Fu \etal~\cite{10.1145/3313831.3376673} created \textit{RestoreVR}. This solution allows visitors to explore, in a VR environment, the mural restoration of the Dunhuang Mogao Grottoes in China. Yi \etal~\cite{8798294} created a VR archaeological excavation system. They propose a real-time immersive excavation simulation that can produce potholes and clods according to the depth and angle of the players' shoveling movements.

\subsection{Education Services} \label{education}

The new generation of students has been immersed in digital technology from birth. Advanced visualization technologies like AR and VR can undoubtedly have a place as pedagogic tools inside and outside the classroom. They can be used to promote collaborative learning and serve as interactive scientific instruments for pupils and teachers.

A tool for building architectural models using VR was created by Raikwar \etal~\cite{8798115}. "\textit{CubeVR}" allows architecture undergraduates to go beyond the limitations of physical models. Plecher \etal~\cite{8797846} developed "\textit{SaMaXVR}", a system that can stream VR content to smartphones. The solution is directed at museums and can be used by visitors wearing disposable cardboard VR headsets. Bäck \etal~\cite{8798101} addressed art education using MR tools. These tools allow the augmentation of facades of historical buildings and iconic works of art with relevant information. The system uses pre-captures 2D images for tracking. Huang \etal~\cite{8446508} evaluated the effectiveness of VR as a learning task for students. They focused their study on using HMDs for learning collaborative automotive assembly tasks. Chiou \etal~\cite{8798339} created a collaborative MR tool for learning how tornadoes are formed. Wearing see-through headsets, students and their instructors can collaborate in the scope of a shared visualization.

Sun \etal~\cite{8798129} compared how students performed in knowledge tests with traditional means and when using a VR environment. The VR version of the tests led to increased student involvement and engagement. Slavova \etal~\cite{8446486} performed a comparative study of VR's learning outcomes and experience in education. She observed increased social interaction and productivity by comparing VR learning with a conventional lecture with slides. Kim \etal~\cite{8798106} developed a learning management system for VR education. The "\textit{VR-MOOC}" tool is directed at supervising students learning in a VR environment. The system allows the students, wearing a VR headset, to communicate with the teacher when necessary. The teacher can monitor every interaction of the students and can intervene directly or indirectly. Lin \etal~\cite{8798331} devised a method to stabilize mentees' first-person video in AR telementoring. Luqrirr \etal~\cite{8446312} compared teacher training in VR with traditional approaches based on video analysis and reflections. Yoshimura \etal~\cite{8798327} created an attention-restoring method for educational VR. They created a gaze-tracking-based system to detect when students' focus shifts away from critical objects in the educational VR environment. Borst \etal~\cite{8448286} analyzed teacher-guided educational VR in the scope of a virtual field trip. The group of students is guided by a live teacher whose image is captured and live-streamed using depth camera imagery. The live image of the teacher is artificially integrated into the VR environment. In a similar framework, Woodworth \etal~\cite{8798318} compared teacher avatar appearances in educational VR. They conclude that students prefer avatars generated using depth camera imagery.

Schiavi \etal~\cite{8446412} created an AR application for teaching history in middle school. The application runs on a tablet where students can point to specific photos of their history manual. Pittman \etal~\cite{8797908} worked on determining design requirements for AR physics education applications. In the scope of VR, Scavarelli \etal~\cite{8798100} point out universal design accessibility and social cognition as determining factors when integrating VR tools in education. On the other hand, Southgate~\cite{8797841} noted how immersive technologies can encompass higher-order cognitive learning skills and be used for self-directed, collaborative, and imaginative learning.

\subsection{Other Applications} \label{other}

As we have seen throughout the previous sections, the application of XR technologies has been researched widely in many areas of science and industry. However, in our survey, other applications did not fall under the areas addressed or cut across multiple domains.

Gonzaga \etal~\cite{8446511} devised a system to take advantage of the detailed field data that the Oil\&Gas industry typically collects. An immersive environment is generated that allows the virtual exploration of petroleum reservoirs. The system incorporates tools that enable detailed geological analysis. Tanaka \etal~\cite{7892374} developed an immersive VR training tool for substation electricians. The system provides a safe environment for trainees to interact with electric equipment, explore the facility, and practice complex maneuvers to recover substation operations. Fritz \etal~\cite{8446324} created a VR data analysis tool for correlating bird populations at a wildlife preserve with pollution. Cho \etal~\cite{7892322} devised a real-time interactive AR solution for broadcasting. The system can perceive the indoor space using broadcasting and RGB-D cameras. It also supports real-time interaction between the augmented virtual content and the casts. Takezawa \etal~\cite{8797923} developed an AR setup for the automotive production industry that allows interior designers to compare material samples. The system uses projection mapping to alter the appearance of the surface of mockups of the car interior.

Regarding the application of XR to improve workplace functionality, Zielasko \etal~\cite{8798068, 8797900} created an interactive virtual desk optimized for visual data analysis tasks. Mori \etal~\cite{7892370} experimented with the application of diminished reality to visualize the work area in an office desk. Grubert \etal~\cite{8446250} tested distinct hand representations for typing in a VR environment. Otte \etal~\cite{8797740} addressed using touch-sensitive physical keyboards for text entry in VR. Zielasko \etal~\cite{8797837} analyzed the use of mid-air menus with passive haptic feedback in traditional office desks.

The experimentation of different professional settings by first-time job seekers in VR environments was addressed by Fominykh \etal~\cite{8798179}. This VR simulation can help get insights regarding preferred workplaces through an immersive and interactive experience. Mohr \etal~\cite{10.1145/2702123.2702490} developed a system that can automatically transfer printed technical documentation, such as handbooks, to a three-dimensional format that can be used in AR. A VR tool for training social skills was created by Taupiac \etal~\cite{8798317}. This tool is directed at managers and
sales representatives and allows role-playing sessions with virtual characters. Moulec \etal~\cite{10.1145/2993369.2993410} studied take-over control in collaborative virtual environments for training. The take-over is the transfer of the interaction control from one user to the other. Tahsiri \etal~\cite{8446431} developed a multisensory virtual environment for Occupational Safety and Health training. The system incorporates heat and olfactory feedback to improve the sense of presence. Knopp \etal~\cite{8446614} addressed the use of industrial robots for providing haptic feedback in VR environments. This solution allows haptic simulation in high-force scenarios and provides a larger feedback amplitude.

\section{Results and Discussion}  \label{discussion}

In the previous section, we explored existing VR, AR, and MR applications research. We focused on how these technologies are being researched for real-world applications. Nevertheless, there is a difference between a solution being usable in a specific professional context and bringing tangible advantages when compared with what is adopted traditionally. In this section, we analyze the different solutions' characteristics, strengths, and disadvantages. We also examine which aspects have made their way to the workflows of companies and institutions. Furthermore, we circumscribe the aspects of XR technologies that determined this applicability.

In Table~\ref{tab:approaches}, we compare the different approaches concerning input/output features. These features include the type of XR technology used (VR, AR, MR, AUV, or a hybrid/multimodal solution). Also shown is the type of display used and input means. Furthermore, we compare the type of feedback that the solution provides. In Table~\ref{tab:sectoral}, we show a sectoral comparison between the approaches. In that scope, we categorize the different solutions by the economic segment in which they fall (sector) and the industry to which they belong. We also describe the primary purpose of the approach. Furthermore, we present examples of real-world applications for some sectors.

\begin{table*}[t!]
\fontsize{8.0}{10.0}\selectfont
  \caption{Input/Output classification of approaches}
  \label{tab:approaches}
  \begin{tabular}{llllll} 
    \toprule
    Reference & Reality & Display & Input & Feedback  \\
    \midrule
			\begin{tabular}{@{}l@{}}\cite{8798295} \cite{8446288} \cite{7892340} \cite{8446127} \\  \cite{8797800} \cite{7223355} \cite{7223463} \cite{7892381} \\ \cite{8798144} \cite{8798249} \cite{8446075} \cite{8797998} \\ \cite{8797918} \cite{7892345} \cite{8798227} \cite{8446160} \\ \cite{8798089} \cite{8797817} \cite{8797997} \cite{8797840} \\ \cite{7223435} \cite{8798239} \cite{8798227} \cite{8798115} \\ \cite{8448286} \cite{7892380} \cite{7892326} \cite{7504785} \\ \cite{8798294} \cite{8446511} \cite{7892374} \cite{8446132} \\ \cite{8446146} \cite{8798041} \end{tabular} & VR & VR Headset & Tracked Object (Active) & AV \\		
	\cite{8797854} \cite{8797761} \cite{7892328} \cite{8798026} & VR & VR Headset & Tracked Object (Active) & AV + Haptic \\
	\cite{10.1145/3290605.3300856} & VR & VR Headset & Tracked Object (Active) & \begin{tabular}{@{}l@{}}AV + Scent\\ + Temperature\end{tabular} \\
	\begin{tabular}{@{}l@{}}\cite{8797872} \cite{8446125} \cite{8797763} \cite{8797828} \\ \cite{8446207} \cite{10.1145/3173574.3173739} \cite{8448290} \cite{7504690} \\ \cite{8446462} \cite{8446370} \cite{7223408} \cite{8798280} \end{tabular} & VR & VR Headset & Tracked Object (Passive) & AV + Haptic \\
    \cite{7892324} \cite{8446504} & VR & VR Headset & Tracked Object (Passive) & AV \\ 
    \begin{tabular}{@{}l@{}} \cite{8798253} \cite{8797846} \cite{8798327} \\ \cite{8446226} \cite{7892338} \cite{7223441} \cite{8798041} \\ \cite{10.1145/3290605.3300466} \cite{8798336} \cite{8798309} \cite{8798309} \\ \cite{10.1145/2821592.2821620} \cite{10.1145/2821592.2821619} \end{tabular} & VR & VR Headset & Tracked Body Part & AV \\    
    \begin{tabular}{@{}l@{}} \cite{8446506} \cite{8446506} \cite{8448292} \cite{7223419} \\ \cite{8798052} \cite{8798159}  \end{tabular} & VR & VR Headset & Tracked Body Part & AV + Haptic \\ 
    \cite{8446553} & VR & VR Headset & Tracked Body Part & \begin{tabular}{@{}l@{}}AV + Scent \\ + Haptic + Wind\end{tabular} \\    
    \cite{8797741} \cite{7892328} \cite{10.1145/3359996.3364272} & VR & VR Headset & Joystick/Controller & AV + Haptic \\
	\cite{7892312} \cite{7223437} & VR & VR Headset & Joystick/Controller & AV \\
    \cite{8446068} & VR & CAVE & Tracked Object (Active) & AV + Haptic \\
    \cite{10.1145/2821592.2821607} & VR & CAVE & Tracked Body Part & AV \\
    \cite{7504713} & VR & CAVE & \begin{tabular}{@{}l@{}}Tracked Body Part \\ + Voice recognition \end{tabular} & AV \\
    \cite{7892259} \cite{10.1145/2821592.2821599} & VR & 2D Screen & Tracked Object (Passive) & AV \\
    \cite{10.1145/2993369.2993397} \cite{8798266} & VR & 2D Screen & Tracked Object (Passive) & AV + Haptic \\
    \cite{10.1145/3139131.3139162} \cite{10.1145/3025453.3025566} & VR & 2D Screen & Tracked Object (Active) & AV + Haptic \\  
    \cite{10.1145/2702123.2702344} \cite{7504695} & VR & 2D Screen & Tracked Body Part & AV \\
    \cite{8797889} \cite{8446139} & VR & 2D Screen & Joystick/Controller & AV + Haptic \\
	\begin{tabular}{@{}l@{}}\cite{7223458} \cite{7892371} \cite{7892371} \cite{8446368} \\ \cite{8798081} \cite{8446292} \end{tabular} & AR & Optical HMD & Tracked Body Part & AV \\
	\cite{8446545} & AR & Optical HMD & Tracked Body Part & AV + Haptic \\
    \cite{8446450} \cite{7223338} & AR & Optical HMD & Tracked Object (Passive) & AV + Haptic \\
    \cite{7223449} & AR & Optical HMD & Tracked Object (Passive) & AV \\
	\cite{8446533} \cite{8797908} & AR & Optical HMD & Tracked Object (Active) & AV \\
    \cite{8446143} & AR & 2D Screen & Speech Recognition + Touch Screen & AV \\
    \cite{7223349} & AR & 2D Screen & Tracked Object (Passive) & AV + Haptic \\
    \cite{8446412} & AR & 2D Screen & Touch Screen & AV \\
    \cite{7504725} & AR & HUD & Tracked Object (Passive) & AV \\  
    \cite{8797923} & AR & 3D Projection & - & AV + Haptic \\
	\cite{8798180} \cite{8798339} & MR & Optical HMD & Tracked Object (Active) & AV \\
	\cite{8798101} \cite{8446583} & MR & Optical HMD & Tracked Body Part & AV \\
	\cite{7892298} \cite{8798069} \cite{8797957} & MR & CAVE & Tracked Object (Passive) & AV \\
	\cite{8797784} & MR & 2D Screen & Keyboard+Mouse & AV \\
	\cite{8446240} & MR & 2D Screen & Tracked Object (Passive) & AV \\
	\cite{10.1145/3290605.3300605} & AUV & Optical HMD & Tracked Body Part & AV \\
	\cite{8446141} & hybrid & CAVE + 2D Screen & Touch Screen & AV \\    
    \bottomrule
  \end{tabular}
\end{table*}

\begin{table*}[t!]
\fontsize{8}{10}\selectfont
  \caption{Sectoral classification of approaches}
  \label{tab:sectoral}
  \begin{tabular}{llllll} 
    \toprule
    Reference & Sector & Industry & Purpose & Applications  \\
    \midrule
	\begin{tabular}{@{}l@{}} \cite{10.1145/2702123.2702344} \cite{10.1145/3290605.3300466} \cite{7892338} \cite{8446143} \\ \cite{8446292} \cite{8446553} \cite{8446125} \cite{7504695} \\ \cite{8446132} \cite{8446146} \cite{8797828} \cite{8798081} \\ \cite{8798084} \cite{8797763} \cite{8797997} \cite{8797817} \\\ \cite{8798336} \cite{8798321} \cite{8798032} \cite{8797872} \\ \cite{10.1145/2821592.2821619} \cite{10.1145/2993369.2993397} \end{tabular} & Health Care & Health Care Facilities & Therapy/rehabilitation & \begin{tabular}{@{}l@{}}\textbullet\textit{\href{https://psious.com/virtual-reality-online-therapy/}{Psious VR teletherapy}} \\ \textbullet\textit{\href{https://www.virtuallybetter.com/}{Virtually Better}} \\ \textbullet\textit{\href{https://appliedvr.io/soothevr/}{Applied VR SootheVR}} \\ \textbullet\textit{\href{https://firsthand.com/health-software/}{Firsthand Cool!/Glow!}}\end{tabular} \\

	\begin{tabular}{@{}l@{}} \cite{8446583} \cite{8446504} \cite{8446075} \cite{8446310} \\ \cite{8446462} \cite{7223437} \cite{7223338} \cite{7892259} \\ \cite{8797998} \cite{8798089} \cite{8798249} \cite{10.1145/2821592.2821599} \\ \cite{7892354} \cite{8446160} \cite{8446450} \cite{8797761} \\ \cite{8797840} \cite{8798159} \cite{8797918} \cite{7223441} \\ \cite{7223435} \cite{7892345} \cite{8446370} \cite{8446469} \\ \cite{8797741} \end{tabular} & Health Care & Health Care Facilities & Medical training & \begin{tabular}{@{}l@{}}\textbullet\textit{Reflekt One} medical training\\solutions;\\ \textbullet\textit{Immerse GE Healthcare}\\ radiology solution;\\ \textbullet\textit{ImmersiveTouch/ImmersiveSim} \\ \textbullet\textit{Medical Realities} \\ \textbullet\textit{\href{https://oxfordmedicalsimulation.com}{Oxford Medical Simulation}}  \end{tabular}  \\
 
	\begin{tabular}{@{}l@{}} \cite{8446488} \cite{8446194} \cite{8798239} \cite{8797858} \\ \cite{8798266}  \cite{10.1145/3290605.3300605} \cite{10.1145/2993369.2993383} \end{tabular} & Health Care & Health Care Facilities & Accessibility design \\
	
	\cite{7223349} \cite{8797763} & Health Care & Health Care Facilities & Surgery planning & \begin{tabular}{@{}l@{}} \textbullet\textit{ImmersiveView IVSP};\\\textbullet\textit{Osso VR}  \end{tabular} \\
	
	\cite{8446502} & Health Care & Health Care Facilities & Medical trialing \\	
	
	\cite{10.1145/3139131.3139162} & Health Care & Health Care Facilities & Surgery assistance \\
	
	\cite{8446368} \cite{8446395} & Health Care & Health Care Equipment & Therapy/rehabilitation & \textbullet\textit{Applied VR SootheVR} \\
	\cite{8798144} & Health Care & Health Care Equipment & Medical instruments testing \\
	
	\cite{10.1145/3025453.3025566} \cite{7892382} \cite{7892381} & Health Care & Health Care Technology & Medical visualization & \textbullet\textit{ImmersiveView VR} \\
	
	\cite{8798191} & Health Care & Health Care Services & Clinical studies \\


	\begin{tabular}{@{}l@{}} \cite{8446486} \cite{8446508} \cite{8797841} \cite{8798327} \\
	\cite{8446486} \cite{8798339} \cite{8797846} \cite{8798101} \\ 
	\cite{8798100} \cite{8798129} \cite{8797908} \end{tabular} & Consumer Discretionary & Education Services & Teaching aid \\
	
	\cite{8446312} \cite{8446412} & Consumer Discretionary & Education Services & Teacher training \\

	\cite{8448286} \cite{8798318} & Consumer Discretionary & Education Services & Field trip simulation \\
	\cite{8798115} & Consumer Discretionary & Education Services & Modeling simulator \\
	\cite{8798106} & Consumer Discretionary & Education Services & Learning management \\
	\cite{8798331} & Consumer Discretionary & Education Services & Telementoring \\

	
	\cite{7892328} \cite{8798041} \cite{8446073} & Consumer Discretionary & Leisure Products & Baseball training & \textbullet\textit{Boston Red Sox} system \\
	\cite{8798309} & Consumer Discretionary & Leisure Products & Basketball training \\
	\cite{7892371} & Consumer Discretionary & Leisure Products & Jogging assistance \\
	\cite{8446506} & Consumer Discretionary & Leisure Products & Climbing training \\
	\cite{8798026} & Consumer Discretionary & Leisure Products & Golf training \\
	
	\cite{8798227} & Consumer Discretionary & Leisure Products & Sports training \\	
	\cite{10.1145/2821592.2821607} & Consumer Discretionary & Leisure Products & Motor learning \\

	\cite{7504785} \cite{7892326} \cite{7892380} \cite{8798294} & Consumer Discretionary & Leisure Facilities & Archaeological site visiting \\
	
	\cite{8446324} & Consumer Discretionary & Leisure Facilities & Data visualization \\

	\cite{8797923} & Consumer Discretionary & Automobile Manufacturers & Car design \\

	\cite{8446139} \cite{8797889} & Industrials & Aerospace \& Defense & Aerial firefighting training \\
	\cite{7504713} & Industrials & Aerospace \& Defense & De-escalation training \\
	\cite{7892294} & Industrials & Aerospace \& Defense & Communication and navigation \\
	\cite{7892312} & Industrials & Aerospace \& Defense & Tank training \\
	\cite{7892324} & Industrials & Aerospace \& Defense & First responder training & \textbullet\textit{ETC/XVR Simulation} solutions \\	
	
	\cite{8446068} & Industrials & Aerospace \& Defense & Marksmanship training & \textbullet\textit{Asterion VR} \\
	\cite{8798280} \cite{10.1145/3359996.3364268} & Industrials & Aerospace \& Defense & Firefighter training \\
	\cite{8797854} & Industrials & Aerospace \& Defense & Weapon calibration training \\
	
	\cite{8798253} & Industrials & Aerospace \& Defense & Aerospace training \\
	\cite{7223463} & Industrials & Aerospace \& Defense & Space vehicle design \\
	\cite{7223408} & Industrials & Aerospace \& Defense & Crew physical exercise \\

    \bottomrule
  \end{tabular}
\end{table*}

\begin{table*}[t!]
\fontsize{8}{10}\selectfont
  \caption*{Table 2: Sectoral classification of approaches (cont.)}
  \label{tab:sectoral2}
  \begin{tabular}{llllll} 
    \toprule
    Reference & Sector & Industry & Purpose & Applications  \\
    \midrule

	\cite{7504702} \cite{8798180} \cite{8448291} \cite{10.1145/2821592.2821595} & Industrials & Research \& Consulting Services & Architectural design & \textbullet\textit{The Wild} design solutions \\
	
	\cite{7223417} & Industrials & Research \& Consulting Services & Urban design  \\
	\cite{8446288} & Industrials & Research \& Consulting Services & Building design  \\
	\cite{8446141} & Industrials & Research \& Consulting Services & Road design  \\
	
	\cite{7223457} \cite{7892347} & Industrials & Construction \& Engineering & Hydraulic simulation \\
	
	\cite{7892340} \cite{8448292} & Industrials & Construction \& Engineering & Construction safety training  \\

	\cite{8446207} & Industrials & Construction \& Engineering & Road inspection  \\
	\cite{8797784} & Industrials & Construction \& Engineering & Construction inspection \\
	\cite{8798295} & Industrials & Construction \& Engineering & Bridge inspection \\
	\cite{10.1145/3290605.3300856} & Industrials & Construction \& Engineering & Fire protection \\
	\cite{8446545} & Industrials & Construction \& Engineering & Utilities network mapping & \textbullet\textit{Augview} \\
	\cite{7892374} & Industrials & Construction \& Engineering & Substation electricians training \\

	\cite{7223355} \cite{8446533} & Industrials & Industrial Machinery & Factory Planning \\
	\cite{7223458} \cite{8446127} & Industrials & Industrial Machinery & Manufacturing guidance & \textbullet\textit{PlyMatch}(Anaglyph) \\
	\cite{7223419} & Industrials & Industrial Machinery & Welding Training \\
	\cite{10.1145/2821592.2821620} & Industrials & Industrial Machinery & Robot operation \\
	\cite{8446511} & Industrials & Industrial Machinery & Data visualization \\

	\begin{tabular}{@{}l@{}} \cite{10.1145/3173574.3173739} \cite{7504690} \cite{8448290} \cite{8446240} \\
	\cite{8446226} \end{tabular} & Industrials & Road \& Rail & Driving simulation \\
	
	\cite{7892298} \cite{8447560} \cite{8798069} \cite{8797957} & Industrials & Road \& Rail & Driving assistance \\
	
	\cite{7223449} & Industrials & Airlines & Aircraft piloting assistance & \textbullet\textit{Aero Glass} \\	
	
	\cite{8797800} & Industrials & Marine & Vessel motion compensation & \textbullet\textit{OMS VR} \\
	
	\cite{7892281} \cite{8446490} \cite{8798052} \cite{10.1145/3359996.3364272} & Industrials & Marine & Underwater training \\

	
	\cite{7892370} \cite{8446431} & Industrials & Human Resource \& Employment Services & General Training \\	
	\cite{8798179} & Industrials & Human Resource \& Employment Services & Career guidance \\
	\cite{10.1145/2702123.2702490} & Industrials & Human Resource \& Employment Services & Technical documentation \\
	\cite{7892322} & \begin{tabular}{@{}l@{}}  Communication \\ Services \end{tabular} & Broadcasting & Live broadcasting \\
	
    \bottomrule
  \end{tabular}
\end{table*}

Looking at the type of XR used, most of the solutions presented are VR-oriented. This tendency happens even though the market share of AR is expected to far eclipse the one of VR by 2021, with the former reaching an expected global value of 83 billion USD, in contrast to the 25 billion USD of the latter's~\cite{TechCrunch2017}. One factor that may account for the preference for VR is the still lower field-of-view of consumer AR headsets compared to VR ones. Tracking in augmented environments also tends to be less stable than in VR environments. It is also trickier to implement, as it needs to account, \eg, for fiducial markers or distinctive features in the surrounding environment. Furthermore, AR headsets are, in general, more expensive.

In most of the research works analyzed, HMDs were the preferred output devices in VR and AR/MR. They have advantages in VR because they are more immersive than other alternatives, such as CAVE installations or flat screens. In AR, and unlike tablets, HMDs provide hands-free operation, which is a fundamental feature in a professional setting. Nevertheless, HMDs, in general, also suffer from important disadvantages. They are still heavy and uncomfortable, mainly if used continuously throughout the day. They also tend to cause motion sickness, although this problem has been addressed in many of the analyzed works~\cite{8446240, 8797800, 8798253, 8797889, 8446194, 10.1145/2993369.2993383}. HMDs also tend to be more unhygienic than, for example, CAVE or tablets/smartphones, primarily when used collectively.

Regarding input devices, although AR/VR and game controllers are still widely used, they have been increasingly replaced by depth cameras and wearable motion sensors. The latter are generally less intrusive and allow for more natural interaction. Likewise, walking or running using omnidirectional treadmills is generally a more natural alternative to locomotion in VR environments than other more static techniques~\cite{brunoHipdirectedWalkinginplaceUsing2017}. The fidelity of locomotion in VR is critical, \eg, in firefighting training applications~\cite{7892324} where the primary concern is to subject the user to experiences as close to reality as possible.

Increasing the sense of presence in virtual environments was addressed in multiple VR applications analyzed~\cite{8798280, 8797889, 8446506, 8446431}. To achieve this increase, many solutions include feedback associated with different senses. Although the fidelity of the AV component of the VR simulation is fundamental, the use of haptic feedback can bring an increased sense of "being there" to the simulation. Even more straightforward "passive" haptic solutions, such as using a real ladder synchronized in position with a virtual one in the VR environment, in work-at-height simulators~\cite{8448292}, can bring a substantial gain in the sense of presence.

The use of combined multisensorial stimuli can contribute immensely to the user's immersion in the simulation. In that sense, we can also more effectively influence how the users feels and how they behaves inside the virtual world. A good example is a solution for the fire safety assessment of buildings presented by Shaw \etal~\cite{10.1145/3290605.3300856}. By subjecting the users to additional thermal and olfactory stimuli, they can have a more realistic experience. Therefore, they will behave like they would in the real world. In that sense, they will hear the sound of the evacuation panic, feel the heat from the approaching flames, and smell the burning furniture.

Some of the research work analyzed in this review has resulted in practical applications in the real world. Other papers mention similar commercial products or XR enterprise applications that inspired its development. Below, we address some practical applications related to the reviewed work. In the field of civil engineering, and in particular, in the scope of municipal infrastructure works, the augmentation of the road surface with 3D pipe maps proposed by Côté \etal~\cite{8446545} has been developed commercially by the New Zealand company Augview \footnote{Augview: https://www.augview.net/}. Augview is a mobile asset management application that uses online data to represent urban underground networks superimposed on the road surface in an AR environment. Unlike the research work, which features AR headsets for visualization, the real-world application uses tablets. In manufacturing, the AR tool for composite layup assistance presented by Freeman \etal~\cite{7223458} has also resulted in an effective real-world application. "\textit{PlyMatch}" is commercialized by the British company Anaglyph \footnote{Anaglyph: http://www.anaglyph.co.uk/}, specialized in solutions for the composites industry. PlyMatch can display, in an AR environment, the computer-generated image of the structure and composite ply being assembled, superimposed to the real material.

Similar solutions to the "\textit{VROnSite}" platform, developed by Mossel \etal~\cite{7892324}, are also commercially available. Multiple versions of such VR simulation systems for immersive training of first responders are proposed by both the Environmental Tectonics Corporation (ETC) \footnote{ETC: http://www.trainingfordisastermanagement.com/} and XVR Simulation \footnote{XVR Simulation: https://www.xvrsim.com/en/}. These VR systems can help safety and security professionals train in collaborative scenarios very similar to those they will find in the field. A VR-based baseball batting training solution similar to the one created by Zou \etal~\cite{8798041} has been used by the North-American baseball team Boston Red Sox to find new promising young batters \footnote{The Red Sox Virtual Reality Home Run Challenge: https://www.youtube.com/watch?v=w0ATO8DKtEs}. The player can practice baseball swings in a virtual environment that simulates a traditional batting center using a VR headset and a mockup bat with a VR controller attached.

\section{Conclusions} \label{conclusions}
This work addressed the effective integration of XR technologies in the real world. We examined existing research concerning VR, AR, and MR applications to contribute with an updated view of how these technologies are being used. We also analyzed some examples of applied XR research that has reached the workflows of companies and institutions. Furthermore, we systematized and classified the different solutions so they could be quickly analyzed individually and as a whole. With that objective in mind, we organized the solutions by input and output, addressing, in particular, the display and input technologies and the type of feedback provided. In addition, we framed the different examined works in specific sectors and industries to get a global view of the potential impact of these technologies in the different areas of society and corporate finance, governance, and culture.

XR technologies can provide multisensory immersive experiences and have grown considerably in popularity in recent years. While VR allows users to be separated from reality and immersed in a simulated experience, AR can augment the user experience by superimposing virtual elements to the real world~\cite{verdelhotrindadeExtendedRealitySafety2020}. The benefits of using XR together or as an alternative to traditional visualization and interaction techniques have been extensively demonstrated in scientific studies. This group of technologies has been shown to bring tangible advantages in many areas of society, even in those where conventional visualization techniques have been long-established. Despite the increasing academic production of applied XR-related works, resulting in fascinating solutions, this has yet to echo in a substantial or compelling adoption of XR in industries and institutions. Only some examples exist where XR solutions have crossed academic and real-world barriers. Although it is becoming common for companies and institutions to have XR solutions available, these are rarely incorporated into the companies' workflows for several reasons. In their everyday tasks, the factory worker and the blue-collar worker (and, more importantly, the corporate decision-makers) continue to favor desktop or mobile touch solutions for planning, designing, executing, and monitoring.

It is immediately apparent that VR still dominates over AR and other XR technologies concerning its popularity in research works. We can see that more than three-quarters of the works analyzed contemplate some form of VR solution. One possible explanation for this discrepancy could be VR hardware's slightly more mature state compared to AR. Nevertheless, AR is slowly but steadily growing in popularity. Indeed, an increase in the percentage of works concerning these technologies (when compared to the total number of analyzed XR works) can be observed in recent years. We also noticed such growth tendency in the percentage of papers (from the total each year) that depicted XR applications, especially between 2016 and 2019.

In the analyzed works, we observed that regarding display technologies, HMDs, namely headsets for VR and optical see-through for AR, are much more preferred over other devices. Nevertheless, 2D screens and the more expensive CAVE solutions are still used. Other technologies, like 3D projection and HUD, appear less in this work's papers. In what concerns input technologies, most solutions use tracking for interaction. This input involves tracking objects and body parts in space and local tracking provided by some specialized mechanical device (\eg, ergometer or haptic stylus). Although touch screens, speech recognition, joystick, and game controllers are featured in some of the works, their adoption is marginal compared to tracking. Tracking technology offers the possibility of more intuitive, natural user interfaces, which can better support immersion. The natural interfaces allow people to interact using intuitive actions grounded in real-world, everyday human behavior~\cite{Preim2015}. In addition to the conventional audiovisual feedback, many solutions provide some haptic stimulus. Others offer a more complete set of feedback, including olfactory and thermal stimuli. These constitute a determining factor in increased embodiment and sense of presence~\cite{sousaRemoteProxemics2016}. The latter is the subjective feeling of immersion experienced by the user, that is, the sense of being present in the virtual environment~\cite{10.5555/216164.216189}. In that context, it is clear that many of the analyzed solutions try to take advantage of the effectiveness of XR visualization and interaction technologies to support immersive analytical reasoning and decision-making in real-world tasks. On the other hand, by allowing situated experiences~\cite{marriott2018immersive} (\eg, superimposition of virtual objects to the real world in an augmented environment), it can better connect the user with the object of analysis.

Regarding the sectoral analysis of approaches, health care is, by far, the industry where more XR solutions are being developed. The analyzed works feature multiple VR and AR solutions for patient treatment, therapy, and rehabilitation. On one hand, by immersing patients in relaxing scenarios, VR can be especially useful in dealing with pre or post-operation stress. It can also hinder pain during minor surgeries without general anesthesia. AR can be used very effectively as a rehabilitation aid, as it can be applied, \eg, in the correction of postural or body movement.
On the other hand, XR can be used for medical assistance as a means of in-surgery or diagnostics visualization and guidance. Furthermore, many solutions have been developed for medical and paramedical training. These include specific medical procedures, like the insertion of needles or general interaction with patients and nursing devices in simulated hospital settings. The second biggest consumer of XR technologies is the education services industry. In that scope, XR solutions have been developed for teacher training, in-class aiding, and simulating field trips or specific on-site virtual scenarios. Aerospace and defense, namely military and law enforcement, can also benefit significantly from using immersive visualization and interaction technologies. Solutions have been developed that allow de-escalation training of police squads and communication and equipment operation virtual drills. XR training is also valuable for handling military weapons and vehicles. Furthermore, virtual training has also been employed in firefighting scenarios, allowing firefighters and first responder personnel to execute critical life-saving tasks efficiently.

As previously mentioned, AEC has been on the front line concerning the early adoption of emerging and innovative visualization and interaction technologies. In that sense, it is the ideal growing ground for applying multisensory immersive experiences to analytical reasoning and decision-making. In the analyzed works, XR solutions have been developed for design and consulting services, like architectural and urban design, and for aiding structural and road project development. Furthermore, on-site XR technologies have been applied in construction inspection and infrastructure mapping. Also, the manufacturing industry can benefit significantly from using these immersive technologies. The analyzed works include planning, guidance, and training solutions in industrial settings. In that context, XR is useful in the design of factories, as a manufacturing aid for factory workers, and for industrial tasks training. The road, aerospace, and maritime industries have also been strong proponents of using XR technologies in their processes. Specialized solutions have been developed to aid truck drivers, aircraft pilots, and maritime vessel operators. Moreover, XR has been used in ship design in the aerospace and defense industries, including space vehicle design.
The human resource and employment services industry has also benefited from XR. Solutions have been developed to train human resources personnel and expose job seekers to the experience of different occupational activities in virtual environments. The solutions examined also include applications to the sports industry. Many applications have been developed that allow athletes and trainers to plan better and execute their tasks to improve athletic performance.

We believe that the four major conferences that were part of our analysis (VRST, CHI, SIGGRAPH, and IEEE VR), together with the significantly extended period analyzed (six years, from 2015 to 2020), gave us a reasonably representative scenario of the present academic panorama of XR research. Furthermore, it gave us an extended perception of the evolution of the technological channeling process between the academic and the real corporate world in the last few years.
The interest in developing XR research work with direct application in the real world has been rising. The relevance of AR and VR and its application to the real world will continue to rise in the coming years. Although there are still many issues to resolve concerning hardware, VR and AR systems have reached a point where they are becoming an essential asset for companies.






\phantomsection
\bibliographystyle{unsrt}
\bibliography{bibliografia.bib}



\end{document}